\documentclass{aa}
\usepackage{graphicx}
\input{psfig}

\usepackage{times}

\def\parn{\par\noindent}

\def\ew{equivalent width}
\def\kms{km~s$^{-1}$}
\def\cm2{cm$^{-2}$}
\def\cm3{cm$^{-3}$}
\def\Texc{$T_{exc}$}
\def\CHP{CH$^+$}
\def\CHl{CH~$\lambda$4300}
\def\CHPlb{CH$^{+}~\lambda$4232}
\def\CHPla{CH$^{+}~\lambda$3957}
\def\HI{H~{\sc i}}
\def\H2{H$_2$}

\def\eq#1{{Eq.~\ref{e:#1}}}    % equation reference
\def\EQN#1{\label{e:#1}}        % eqn labelling a la Texsis
\def\Tab#1{{Table~\ref{t:#1}}}        % table reference
\def\Fig#1{{Fig.~\ref{f:#1}}}    % figure reference
\def\units#1#2{#1$^{#2}$}

\begin{document}

\title{Small scale structure in molecular gas from multi-epoch
observations  of HD~34078\thanks{Based on observations made at
Observatoire de Haute Provence (France) and McDonald Observatory.}}

\titlerunning{Small scale structure in molecular gas}

  \author{ Emmanuel~Rollinde \inst{1}, Patrick~Boiss\'e \inst{2,1},
S.R.~Federman \inst{3}, K. Pan \inst{3}}
  \authorrunning{Rollinde et al.}
   \offprints{rollinde@iap.fr}

\institute{$^1$ Institut d'Astrophysique de Paris, 98 bis boulevard
       d'Arago, 75014 Paris, France \\
       $^2$ Radioastronomie, UA CNRS 336, Ecole Normale Sup\'erieure, 24,
rue Lhomond, F-75231 Paris Cedex 05, France \\
        $^3$ Department of Physics and Astronomy, University of Toledo,
Toledo, OH 43606 USA
}

\date{Typeset \today ; Received / Accepted}

\abstract{We present spectroscopic observations of the runaway reddened star
HD~34078 acquired during the last three years at Observatoire de Haute
Provence and McDonald Observatory as well as other spectra obtained
since 1990. The drift of the line of sight through the foreground
cloud due to the large transverse velocity of HD~34078 allows us to 
probe the spatial distribution of CH, \CHP, CN and DIBs carriers 
at scales ranging from about 1 AU up to 150 AU.  In particular, time 
variations in the equivalent width of absorption lines are examined. A 
few past and recent high resolution observations of CH and \CHP\ absorption
are used to search for line profile variations and to convert equivalent 
widths into column densities.

The data set reveals a 20\% increase in CH column density over the past 10 years
with no corresponding variation in the column density of \CHP\ 
or in the strengths of the 5780 and 5797
\AA\ DIBs. CN observations indicate that its excitation temperature
has significantly increased from $<$ 3.1 K  in 
1993 to 3.6 $\pm$ 0.17 K in 1998 while the CN column shows only a
modest rise of $\approx 12 \pm 6$\%. The data also strongly 
suggest  the existence of weak correlated variations in CH and 
\CHP\ columns over periods of 6 - 12 months (or $\approx$ 10 AU).

These results are discussed in relation to \CHP\ production
mechanisms. A dense newly intervening clump is considered in order to
explain the long-term increase in the column density of CH, but such 
a scenario does not account for all observational constraints.  Instead, 
the observations are best described by \CHP\ production in a 
photodissociation region, like that suggested for the Pleiades and IC~348.
\keywords{
{{\it  ISM}:    molecules -   {\it  stars}: individual (HD
34078) - {\it  ISM}: structure  }
}}

\maketitle

\section{Introduction}

While the existence of AU-scale structure is relatively well
established in atomic gas (see e.g., \cite{dieteral76}, \cite{diamondal89}, 
\cite{frailal94}, \cite{lauroeschal99}, \cite{faisonal01}, 
\cite{weltyal01}), the reality of such tiny fluctuations within 
molecular gas is still questionable. 
Indeed, only minor species can be easily observed
towards molecular clouds and their spatial distribution might not
reflect that of \H2\ if ``chemical structure'' were present.
H$_2$CO, HCO$^+$ and OH apparently display column density fluctuations
reaching 5 to 15\% along lines of sight separated by about 10 AU
(\cite{marscheral93}, \cite{mooreal95}, \cite{lisztal00}), 
while dust grains appear to be more smoothly distributed 
(\cite{thoravalal96}). At larger scales (about 10,000 AU), 
Pan et al. (2001) find significant differences in CN, CH and \CHP\ 
absorption lines.

If marked enough, small scale structure within molecular gas might 
notably affect its time evolution, fragmentation and then utimately, 
star formation. It is therefore important to characterise the
properties of such media, identify the parameters displaying
fluctuations and quantify their amplitude and scalelength.  
To this aim, we have selected a bright O9.5 runaway star, HD~34078 
(AE~Aur), seen through a translucent cloud with $E$($B-V$) = 0.52 
(\cite{diplasal94}).
Thanks to its large transverse velocity, about 100 \kms\ for an 
assumed distance of 530 pc, the comparison of spectra taken at one
year intervals provides a measurement of column density variations at 
a scale of 17 AU (assuming a cloud distance of 400 pc: \cite{brownal95}). 
To complement ongoing FUSE observations designed to investigate in a
direct way small scale variations of the \H2\ column density (and
then study {\it density} structure within molecular gas), we have 
undertaken repeated ground-based observations of CH, \CHP and CN 
absorption lines and diffuse interstellar bands (DIBs) towards HD~34078. 

These optical spectra complement the FUSE UV data in several 
important respects:

- since CH is seen to correlate well with H$_2$ over large scales
(e.g., \cite{federman82}), it should be a good indicator of cloud 
structure. On the other hand, the abundance of \CHP\ is largely independent of 
H$_2$ column density, $N$(\H2); the large abundance of this species 
is still poorly understood and hot pockets of gas (e.g., 
shocks or vortices as proposed by Falgarone \& Puget, 1995) could
be the formation site. Then, \CHP\ observations may reveal the 
structure of these presumably very small regions where energy is 
actively dissipated. Finally, CN depends nonlinearly on H$_2$ and 
is thus an indicator of gas density (\cite{federmanal94}).

- given the brightness of HD~34078, high S/N spectra can be 
obtained easily to search for very small fluctuations which would be
difficult to detect in FUSE spectra,

- spectral data in the visible range can be acquired much more easily than
in the far UV, and therefore, it is possible to control the time (i.e., 
spatial) sampling and to accumulate a larger number of spectra.  Moreover,
several visible observations of HD~34078 have been performed in the past 10
years, thus giving access to large scale variations. A broad range
of scales can then be explored through visible observations.

- high resolution observations providing information on the velocity
distribution and its possible variations are feasible in the visible, but
not in the FUSE range.

In this paper, we present observations made specifically for this project
together with a comparison with older data. The whole set of spectra is
used to search for time variable absorption and infer the structure
implied for CH, CH$^+$, CN and diffuse bands carriers.
Sect. 2 presents the observations and the methods used to analyse
the data. In Sect. 3, we give the results obtained for CH, \CHP, CN
and some selected DIBs. Since most observations have been obtained at 
low resolution, we mainly consider equivalent width variations. 
High resolution observations of CH and CH$^+$ are then used to infer 
velocity distributions and translate 
variations in equivalent width into column 
density variations for these two species. The implication of our 
results for the existence of AU-scale structure in the distribution of 
these tracers is discussed in Sect. 4.

The analysis of Cycle 1 and Cycle 2 FUSE data will be presented in
a separate paper (Boiss\'e et al., 2002; see also Boiss\'e et al., 2001 
and Le Petit et al., 2001 for a preliminary report). Our ultimate goal 
is to correlate the variations seen for \H2\ and those of other tracers 
like CH, \CHP\ and CN in order to improve the description of the structure.

\section{The Data}
\label{s:data}
\subsection{The Available Spectra}

\parn 
Spectra acquired at various observatories have been used for this
study.  In the following, these are named according to the observer 
and year of data taking. We obtained specifically for this 
project a series of nine spectra at the Observatoire de Haute 
Provence (OHP) ({\bf R99-02}). In addition, high
resolution spectra were recently recorded at McDonald 
Observatory ({\bf F02}). The motivation was twofold. First, as we 
are interested in {\it column density} variations, we need to translate 
the observed changes in 
equivalent width, $W_{\lambda}$, into changes in $N$ (i.e., determine $N$ and
d$N$/d$W_{\lambda}$).  Since the lines are not optically thin, 
the conversion requires a good
knowledge of the velocity distribution. This point is especially important
for \CHl\ because this feature is in fact a blend of two transitions and
the analysis can be notably affected by the splitting if the $b$ parameter is
less than 2 \kms\ (\cite{lien84}). Second, only high resolution
observations allow us to determine whether or not line shapes have
changed, and they possibly reveal profile variations not associated with  
equivalent width changes.

HD~34078 was observed several times since 1991 by one of us
(S. Federman at McDonald Observatory: {\bf F93a, F93b}). 
Spectra of HD~34078 were also acquired by M. Allen (KPNO: {\bf A91}),
P. Jenniskens and F.X. D\'esert (at OHP: {\bf J91}), 
J. Krelowski, G. A. Galazutdinov \& F. A. Musaev (at Terskol Observatory: 
{\bf K97a, K97b}) and G. Herbig (at
the Keck telescope: {\bf H98}) who kindly provided their data, 
so that we could analyze them in the same way as our own and 
search for variations by direct comparison of the spectra. 
Note that in his study, Herbig (1999) searched for changes in line
{\it position} with respect to the spectrum obtained by Adams (1949)
and thus investigated the {\it velocity structure} of the foreground
gas; no significant variation was observed.
 
\begin{figure}
	\unitlength=1cm
	\begin{picture}(10,8)
        \put(0,0){\psfig{width=8cm,figure=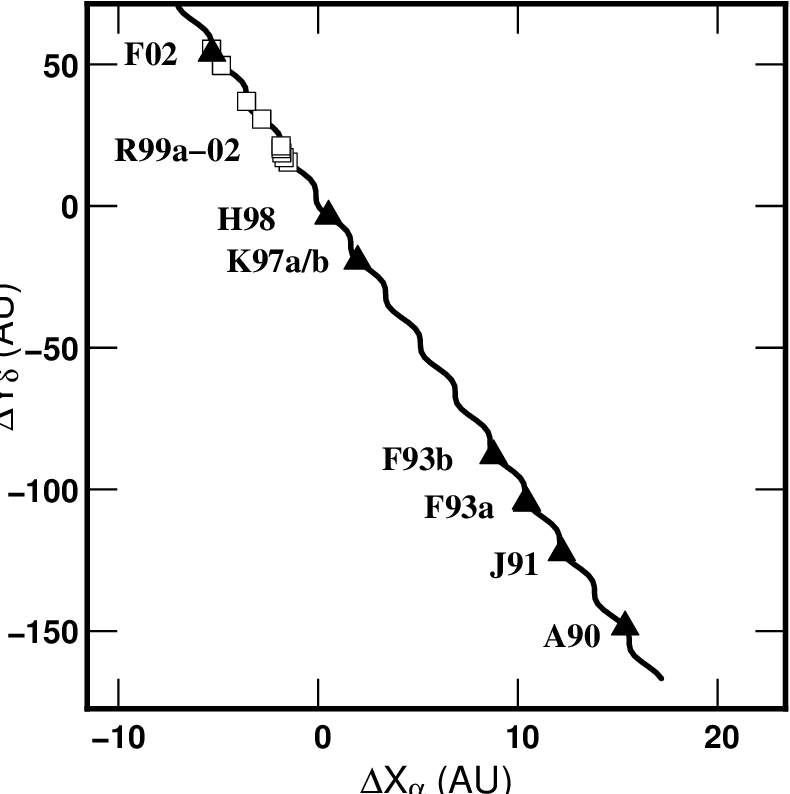}}
        \put(4.5,4.5){\psfig{width=3cm,figure=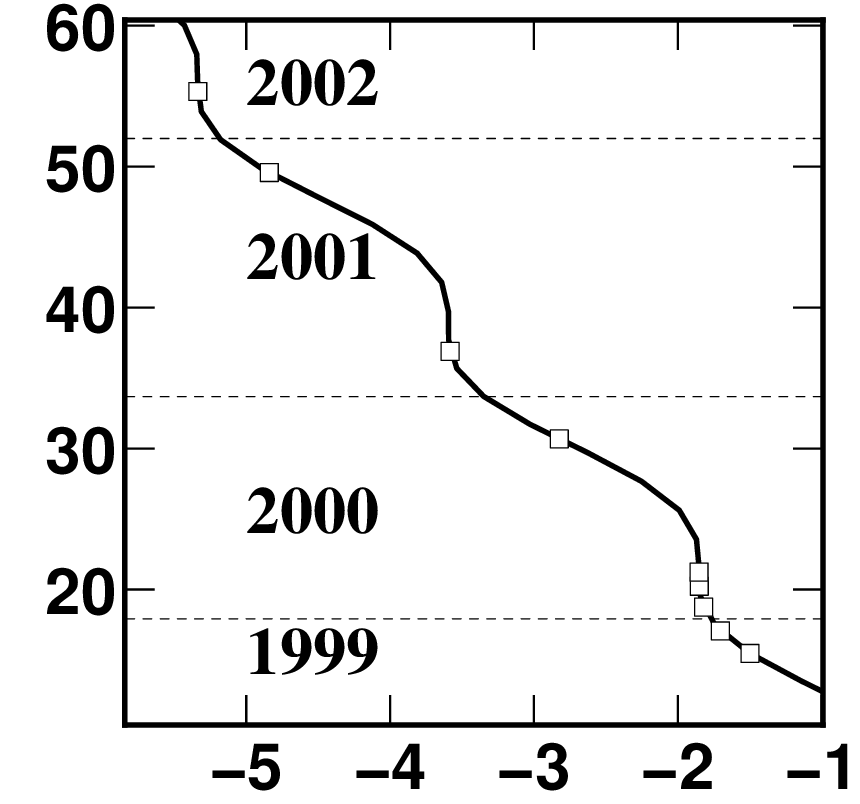}}
	\end{picture}
\caption{The impact point of the line of sight towards HD~34078 onto
the foreground cloud versus time. The motions of the earth, the sun
and HD~34078 have been taken into account. Markers specify the position
of the star for each epoch at which spectra were obtained
(cf. Table~\ref{t:observation}). Note that the X and Y scales are not
identical. The insert focuses on the recent observations made for this
project ({\bf R99a-R02, F02}) that probe the smallest scales.}
\label{f:position}
\end{figure}

Table~\ref{t:observation} summarises all these observations
and gives the date, resolution and signal-to-noise ratio. 
The transitions (from CH, \CHP, CN or DIBs) contained in the 
observed wavelength range are indicated. These data altogether 
provide a sort of 1D cut through the
cloud, the sampling of which is illustrated by \Fig{position}.
Below, we give some details concerning the observations of HD~34078
used in this paper and not previously published.

\subsection{OHP Observations}
\begin{figure}
\centerline{\includegraphics[width=6cm,angle=-90]{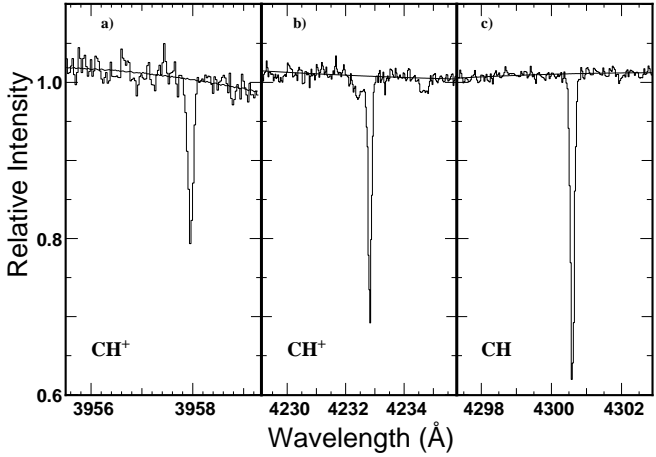}}
\caption{The \CHP\ lines at 3957 and 4232 \AA\ (a,b) and 
CH line at 4300 \AA\ (c) in the
OHP spectrum recorded in December 1999. The raw spectrum has just been
divided by a constant to scale the mean intensity to 1. Note the broad 
Fe~{\sc II} stellar feature blended with the 4232 \AA\ line. The S/N is 230 
per 0.03 \AA\ pixel around 4300 \AA.}
\label{f:observation}
\end{figure}

The spectra were obtained using the fiber-fed echelle spectrograph
Elodie (\cite{baranneal96}) mounted on the 1.93 m telescope 
({\bf R99-02}). The resolution is about 32000.
Initially, a spectrum was taken each month in order to probe very small
scales. Since this series did not reveal significant variations, only two
spectra were taken during the next winters. The range covered  is 3906 \AA\ -
6811 \AA\ and includes lines from CH (4300 \AA), CH$^+$ (3957
and 4232 \AA) and DIBs. For all spectra, the integration time is about
1 hour, split in 2 to 3 successive exposures of 30 or 20 minutes each 
in order to check the stability of the instrument and robustness of
our analysis on independent spectra. The corresponding S/N ratio is 
typically 150 per resolution element. Some examples of 
CH and \CHP\ lines detected in OHP spectra are shown in \Fig{observation}.
The spectra have been extracted and wavelength calibrated using the 
automatic on-line data reduction program attached to the instrument
(see \cite{baranneal96} for details); the wavelength calibration is
based on Thorium lamp spectra.

\subsection{McDonald Observations}

Three data sets were acquired since the last published results (Federman
et al., 1994) by one of us (S.R.F.).  Two of the new sets utilized the
6 foot camera and the 2dcoude spectrograph on the 2.7 m telescope.  A
standard observing strategy, with bias frames, flat fields, and a Th-Ar
comparison spectrum, was employed (see Knauth et al., 2001 for details of
similar measurements).  Individual orders containing CH$^+$ $\lambda$4232
and CH $\lambda$4300 were imaged with the 6 foot camera in 1993.
The nominal resolution was 1.5 km~s$^{-1}$.  In 2002, spectra
of CN, CH and CH$^+$ were acquired at a single setting
of the 2dcoude spectrograph.  Here, the resolution was about 1.9 km~s$^{-1}$. 
 Also in 1993, the Sandiford echelle spectrograph on the 2.1 m
telescope (McCarthy et al., 1993) was used for measurements on various
species in a similar fashion.  One setting, centered on 3990 \AA, provided
data on CN, CH $\lambda\lambda$3878,3886,3890 and 
CH$^+$ $\lambda$3957.  The second setup at 4300 \AA\
measured absorption from CH$^+$ $\lambda$4232 and CH $\lambda$4300.
The widths of Th-Ar lines indicated a resolution of $\sim$ 7.5 km~s$^{-1}$
for these spectra.  All data were reduced and analyzed with the IRAF
package.

\subsection{Analysis of the Data}
\label{s:analysis}
\parn 
In most of the available HD~34078 spectra, the profiles are
not resolved. We shall then look primarily for variations in $W_{\lambda}$. 

We have been particularly careful in placing the
continuum in a consistent manner for all epochs and also in 
estimating properly the uncertainty in $W_{\lambda}$.
For some species, several lines are observed, e.g., at 3957 \AA\ and
4232 \AA\ for \CHP.  Similarly, a few observations of the CH lines at 
3878, 3886 and 3890 \AA\ are available for comparison with the
4300 \AA\ results. This redundancy is useful as a means
to assess the robustness of our analysis and validate
the error estimates.

We have developed an automatic procedure based on the MIDAS 
package {\it alice} in order to fit 
the continuum in a user-independent way and to measure $W_{\lambda}$ values.
Several ``clean'' windows (generally two, located on both sides of the line)
 are selected; a polynomial fit with a degree ranging from
1 to 3 and depending on the shape of the spectrum is then performed.  
This provides a good estimate of the continuum in the interval covered 
by the absorption feature (see \Fig{observation}).
One $\sigma$ uncertainties in $W_{\lambda}$ can be estimated as the quadratic sum of
$\epsilon_{n}$ and $\epsilon_{c}$, the pixel-to-pixel noise and
the uncertainty in the continuum level in the normalised spectrum
respectively. The latter are given by

\begin{eqnarray}
  \epsilon_{n}&=&\sigma_{F} \times \delta\lambda \times \sqrt{N}\;\,,
{\rm\ \  
and}\label{e:err_noise}\\
  \epsilon_{c}&=& \Delta\lambda_{line} \times \mid 1- F \mid
\;,\EQN{err_sys}
\end{eqnarray}
where $\Delta\lambda_{line}$ is the total width of the absorption
feature, $\delta\lambda$ is the pixel size and $N$ the number of
pixels covered by the line  ($\Delta\lambda_{line}=N\delta\lambda$).
The average (\eq{err_sys}) and the r.m.s. (\eq{err_noise}) of the normalised
flux, $F$, 
are computed along the windows adjacent to the line.
The robustness of the fitting procedure and the estimates for uncertainty 
have also been checked against synthetic spectra including absorption
lines with known $W_{\lambda}$, a signal-to-noise ratio comparable to that in
our data, and various continuum shapes. 

Some transitions present difficulties. \CHP$\lambda$4232 is blended 
with a shallow stellar feature (from Fe~{\sc II}, see \Fig{observation}b) 
and the above procedure cannnot be applied. Similar problems are 
encountered with the R(0) CN line at 3876 \AA. In such cases, the $W_{\lambda}$ is 
measured using the command {\it integrate/line}
from MIDAS (an order 2-3 is chosen for the polynomial used to fit the
spectrum immediately adjacent to the line) and the uncertainties are
determined from successive measurements.

The high resolution profiles have been fitted using VPFIT 
(\cite{carswellal87}). No attempt was made to look for velocity shifts 
(the question addressed by Herbig, 1999).

\section{Results}
\label{s:result}

We now present the results obtained for CH, CH$^+$, CN and some
selected DIBs. For CH and CH$^+$, we first discuss the homogeneous set of data
obtained after November 1999 which probe time scales ranging from one
month to 2.5 years (corresponding to 1.5 - 40 AU). The whole set of
available observations made in the past 11 years are presented
next. Finally, we discuss the high resolution data and their
implications for velocity structure and {\it column density} variations.

\subsection{Structure in CH}
\label{s:CH}
 
\subsubsection{Small scale variations}

\begin{figure}
	\unitlength=1cm
	\begin{picture}(10,20)
        \put(0.7,15){a)}
        \put(-0.1,14){\psfig{width=8.5cm,angle=-90,figure=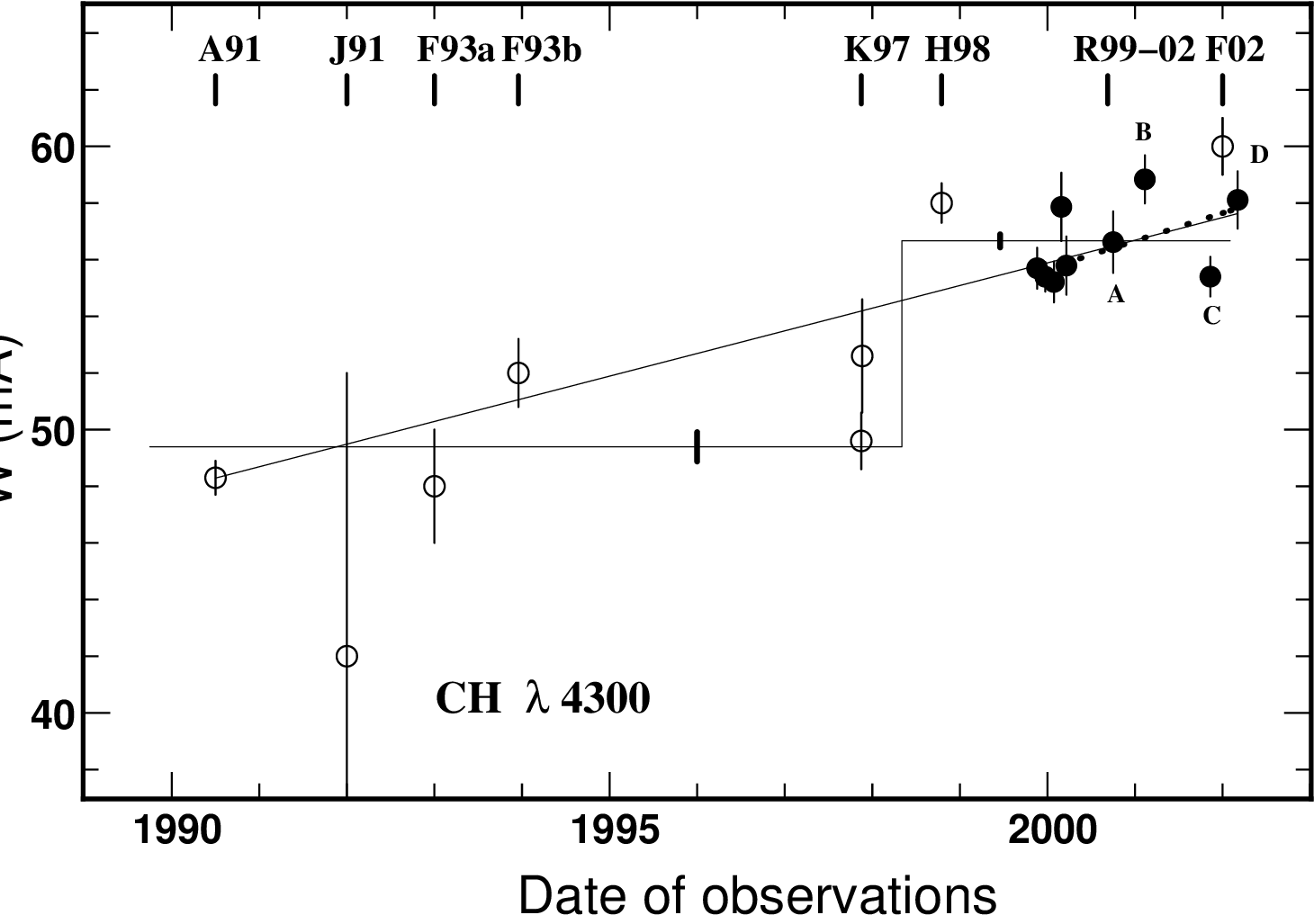}}
        \put(0.7,8){b)}
        \put(-0.1,7){\psfig{width=8.5cm,angle=-90,figure=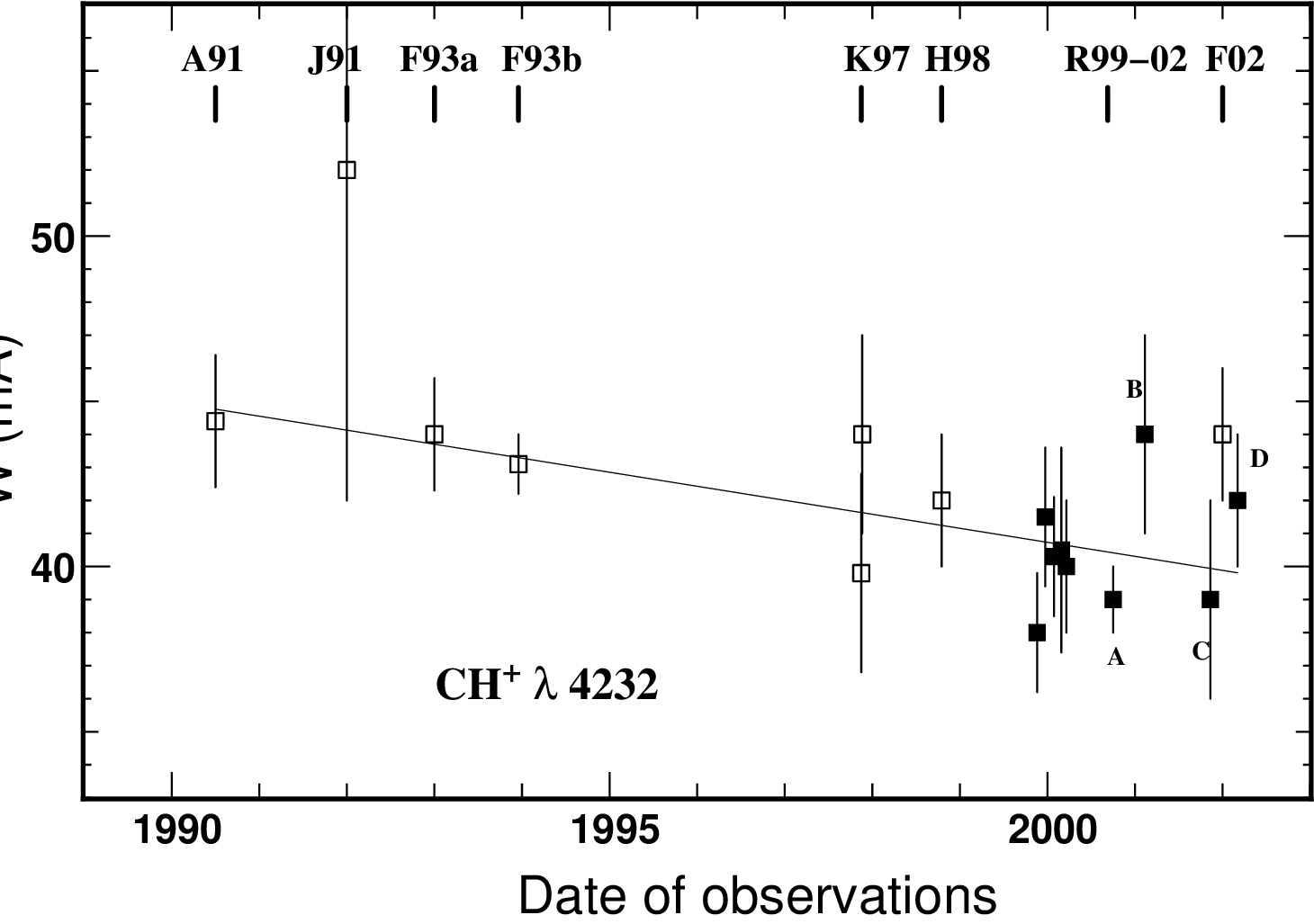}}
        \put(0.7,1){c)}
        \put(-0.1,0){\psfig{width=8.5cm,angle=-90,figure=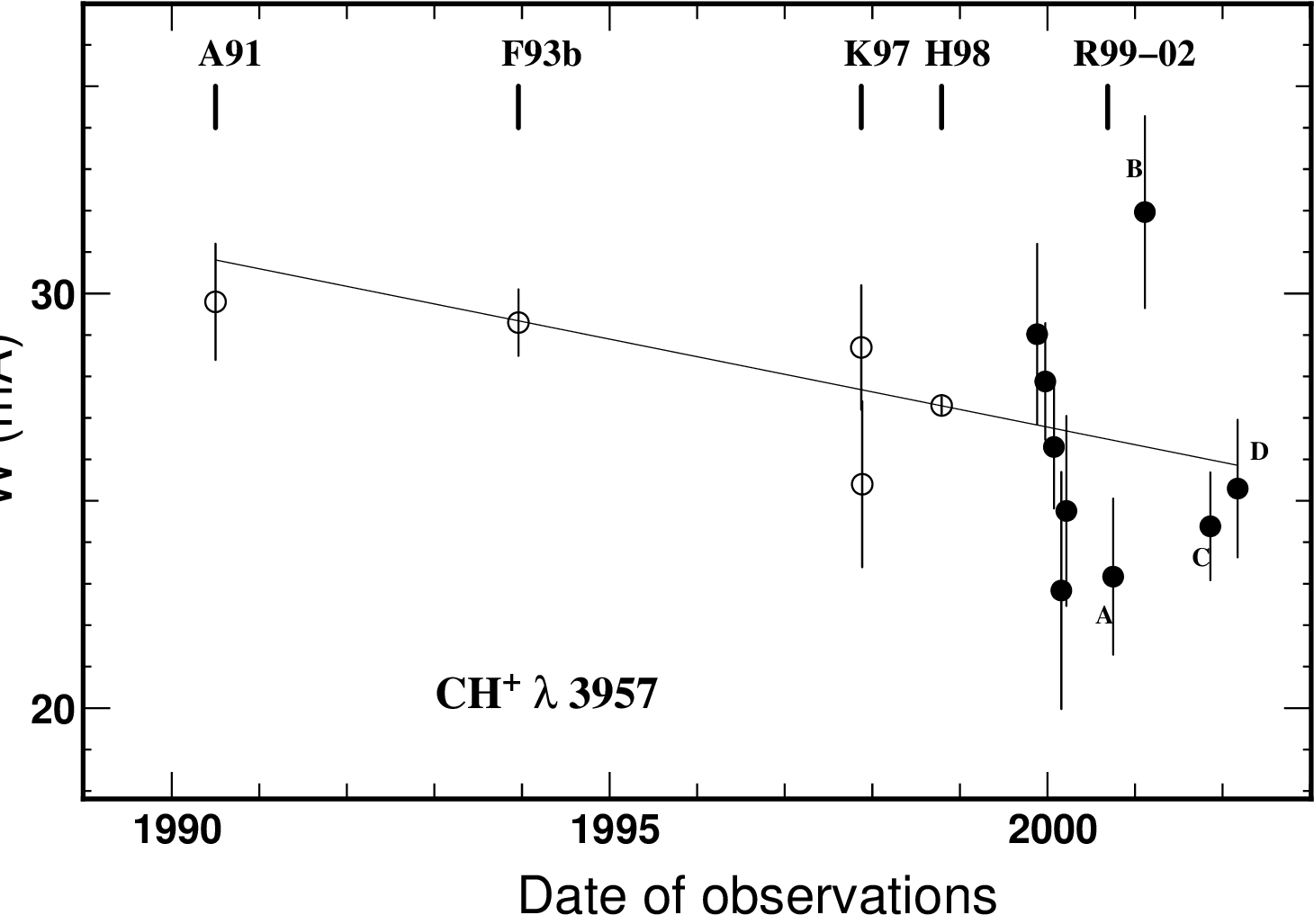}}
	\end{picture}
\caption{The equivalent width of CH and CH$^+$ lines versus time. 
{\em Top panel}: \CHl; {\em middle panel}: \CHPlb. 
{\em bottom panel}: \CHPla. Solid markers show recent
closely-spaced OHP observations. A linear
fit is performed on the whole data set 
(solid lines). For recent OHP/F02 data on \CHl, a separate 
linear fit is given (dotted line on top panel). A fit by a step
function is also shown for the CH data, with errors for 
the upper and lower values indicated by the small vertical bars. 
Note the similarity between CH and CH$^+$ variations from October
2000 to March 2002 (points labelled A, B, C and D), suggesting the
reality of correlated erratic fluctuations.  Such fluctuations are 
also suggested by the McDonald data of {\bf F02}.}
\label{f:result_ohp}
\end{figure}
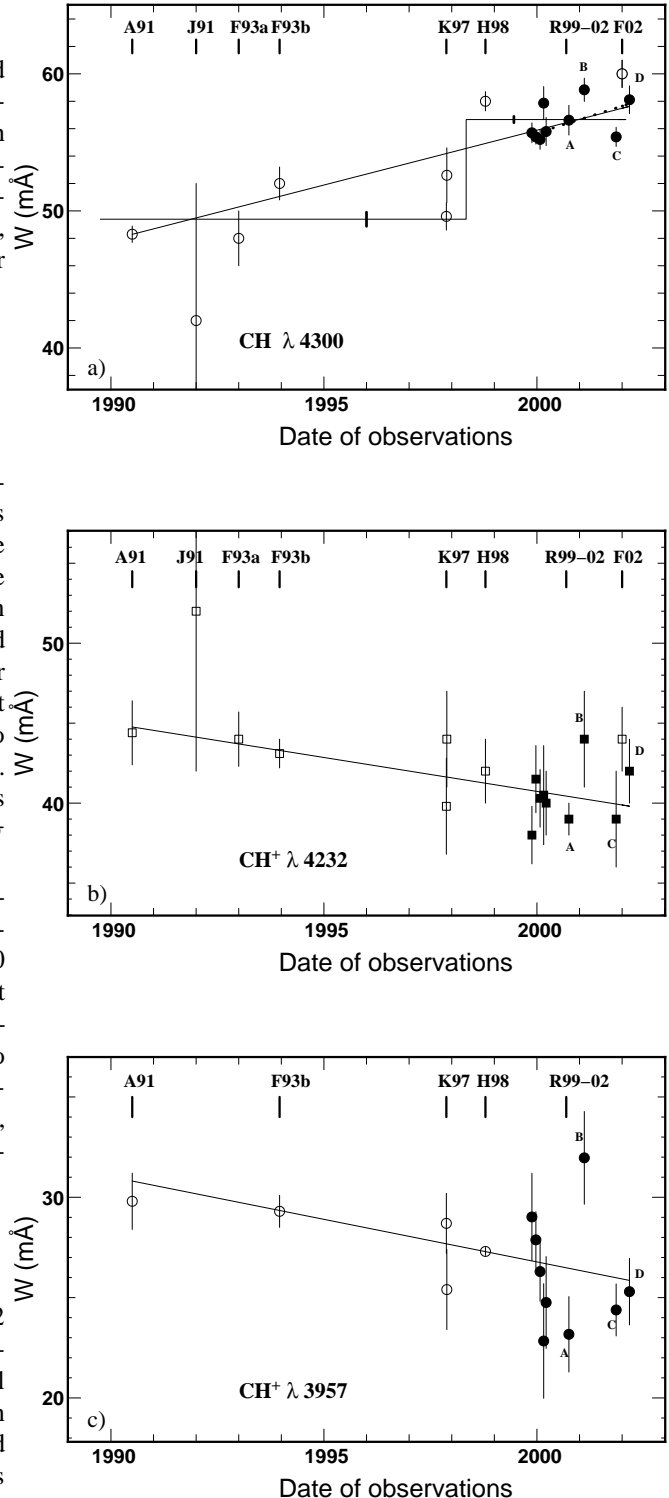

\Fig{result_ohp}a displays the evolution of the
\ew\ as a function of time for the 4300 \AA\ line.
The variation since 1999 is fitted with a linear 
form $W_{\lambda} = W_{\lambda,0} + \alpha\, (t-t_{\rm med})$ where 
$t_{\rm med}$ is the median of observing times (December 2000). 
The data points are weighted according to signal-to-noise ratio in all
fits performed in this paper. 
The contours of the reduced $\chi^2(W_0,\alpha)$ are plotted in
\Fig{chi}a (dashed lines) for the linear case.  
A constant value is rejected only at the 1$\sigma$ level. 
The best fit,  $W_0 = 56.6$ m\AA\ and $\alpha = 0.98$ m\AA\ yr$^{-1}$ 
corresponds to $\Delta W_{\lambda}\ \simeq $ 2.2
m\AA\ between November 1999 and February 2002.
It is overplotted in \Fig{result_ohp}a (dotted line).
Large variations at this scale are excluded since 
$-1.8 < \Delta W_{\lambda}\ < 6.3$\ m\AA\ at
the 3$\sigma$ level.

The linear fit is not fully satisfactory since  residuals correspond to
$2\sigma$.
This suggests that small additional erratic variations in $W_{\lambda}$ may be
present. The observed amplitude is 1.50 m\AA\ ($rms$) about the linear fit
(and 1.56 m\AA\ about a constant value; the maximum difference is 4.8
m\AA). A Kolmogorov-Smirnov test (Miller, 1956) 
shows that these values are too small compared to 
observational errors to indicate with certainty that we are detecting 
fluctuations in $W_{\lambda}$(CH). However, the reality of the latter
is supported by the fact that similar
variations are observed for \CHP\ lines (see below).

\subsubsection{Large scale variations}
\begin{figure}
\centerline{\includegraphics[width=8cm]{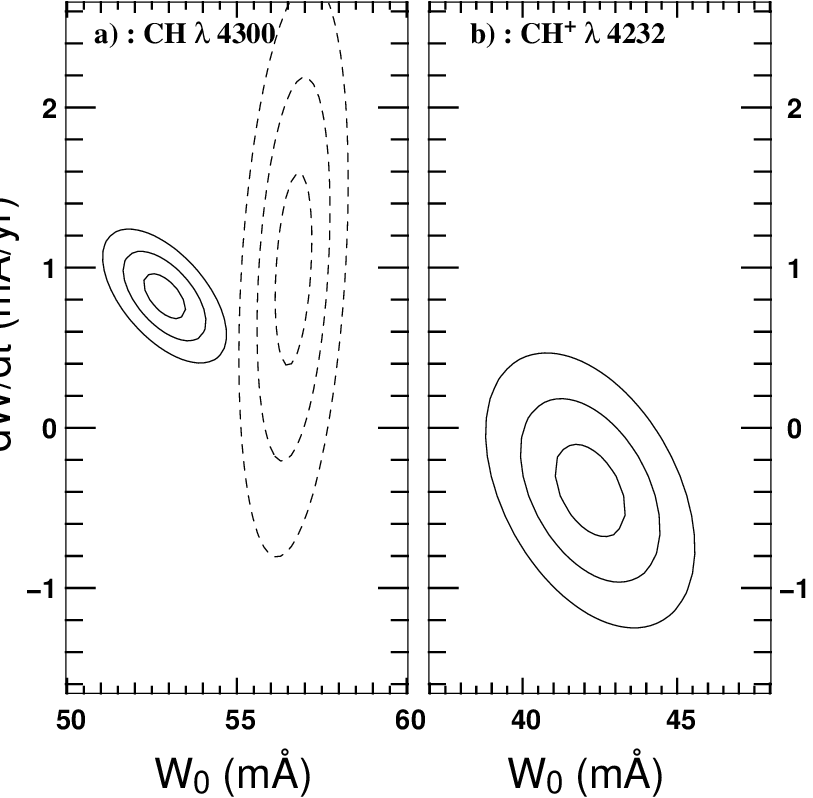}}
 \caption{Linear fit of $W_{\lambda}(t)$ for \CHl\ (a) and 
\CHPlb\ (b) ($W_0=W_{\lambda}(t_{\rm med})$). 
  {\em Solid line}:
 linear fit to the whole set of data (with $t_{\rm med}$ = April 1996).
 {\em Dashed line}: linear fit 
 to the recent data ($t_{\rm med}$ = December 2000). The best fits are shown in \Fig{result_ohp}.
 Contours represent the levels 1, 2 and 3 of the reduced 
($\chi^2-\chi^2_{\rm min}$).}
 \label{f:chi}
\end{figure}

Results for the full set of observations are given
in \Tab{result} and shown also in \Fig{result_ohp}a. 
A significant increase is apparent. Fitting again with a linear variation 
(with $t_{\rm med}$ = April 1996) 
gives a total increase of $\Delta W_{\lambda}$ = 8 m\AA\ or 16\% between 1992 and 
November 2001, consistent with the slope derived from the small scale 
data alone (see \Fig{chi}). The 3$\sigma$ limit is 
$4.2<\Delta W_{\lambda}<13$ m\AA.
However, the fit is even less satisfactory than for the recent 
data alone since the minimum $\chi^2$ is larger than $3\ \sigma$.
The assumption of a linear increase is somewhat arbitrary and visual
inspection of \Fig{result_ohp} suggests that a step function might
also provide an acceptable fit. The minimum $\chi^2$ is not 
significantly 
lowered, $\sim 2.6\sigma$. The date at which the step 
occurs is well constrained by {\bf K97} and {\bf H98} measurements. 
The amplitude ($\Delta W_{\lambda} \sim 7$\ m\AA) is similar to the total 
increase inferred from the linear fit. To assess the reality of an
increase in $W_{\lambda}$(4300) regardless of any assumption on its form, we 
 performed a Pearson test to investigate the correlation between 
$W_{\lambda}$ and time. An uncorrelated evolution of \CHl\ with time is 
rejected at the 4~$\sigma$ level.

Other CH lines at 3886 and 3890 \AA\ have been observed in 1990, 1993
and 1998.  Their variation should be consistent with the increase seen
in \CHl. Given the weakness of these lines and the S/N attained, we
find that even the strongest feature at 
3886 \AA\ does not bring useful additional constraints.

\subsubsection{High resolution observations}

\begin{figure}
 \centerline{\includegraphics[width=6cm,angle=-90]{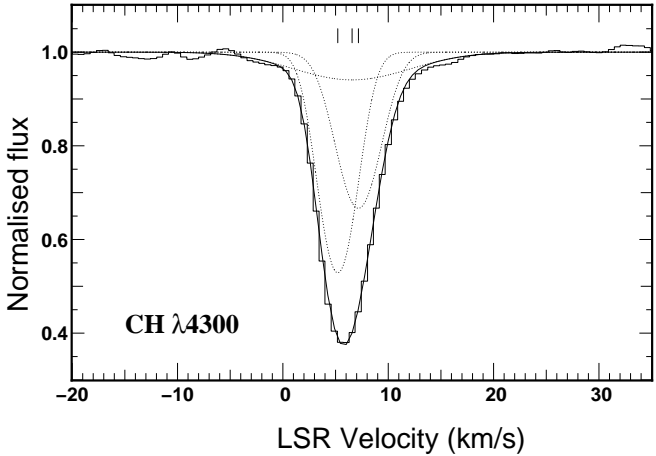}}
 \caption{High resolution observation ({\bf F02}) of the CH line.
   {\em solid line}: fit with VPFIT. {\em dotted lines}: the three
   components of the fit.}
\label{f:profCH}
\end{figure}

We first discuss the highest signal-to-noise ratio observation of \CHl,
performed in 2002 at McDonald ({\bf F02}). The profile was fitted
using VPFIT, taking into account the doublet structure of the 
CH ground level (Lien, 1984). Both levels are assumed to have the 
same column density. 
Three components are needed: two narrow ones reproduce
the slightly asymmetric core of the absorption line and a much broader and
fainter one accounts for the wings which are present on both sides of
the core. The resulting fit and the three components are shown in 
\Fig{profCH} and the parameters of the fit are given in \Tab{vpfit}.
There is no doubt about the reality of the broad component since it is
clearly seen in all spectra with the appropriate S/N ratio ({\bf A91},
{\bf F93a}, {\bf F02}) and for both the 3886 and 4300 \AA\ lines 
({\bf H98}). It is noteworthy that, although barely noticeable, it contains 
no less than 28\% of the total CH column density. As a consistency
check, we verified that $W_{\lambda}$ values measured for all CH lines 
detected in Herbig's spectrum are consistent with the velocity
distribution derived from the {\bf F02} spectrum. 
We note, however, that different 3-component solutions 
for the latter are also acceptable (in particular, one in 
which the strongest narrow component is the red one).
Nevertheless, since the optical depth remains moderate, this
degeneracy does not result in larger uncertainties on $N$(CH). 
Moreover, very recent observations of the CO(2 - 1) emission
line at 1.3 mm towards HD~34078 performed at the 30 m IRAM telescope give a
velocity profile in excellent agreement with that shown in
\Fig{profCH} (E. Roueff and M. Gerin, private communication).

\begin{figure}[h]
\unitlength=1cm
\begin{picture}(10,7.5)
\put(0,-1.2){\psfig{width=8cm,figure=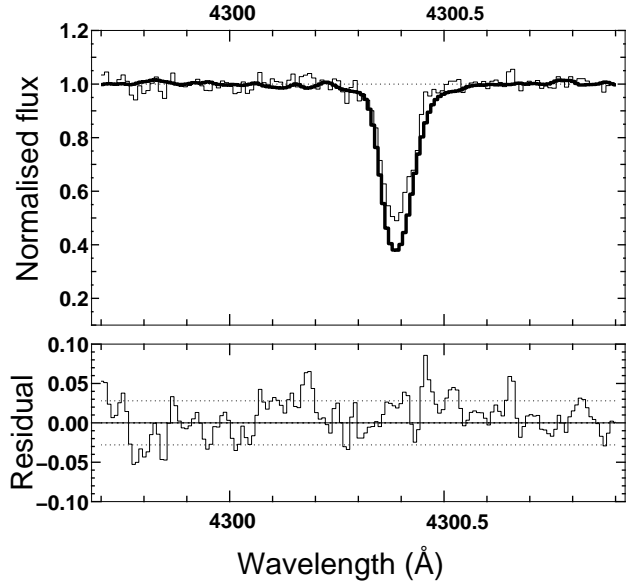}}
\end{picture}
\caption{Variation of the CH profile. {\em Top panel}:
 {\bf F93a} and {\bf F02} spectra are overplotted as thick and solid line 
respectively. {\em Bottom panel}: the optical depth of {\bf F93a} 
is multiplied by 1.34 and the resulting spectrum is subtracted from {\bf
F02}. The residual (solid line) is compared to the 1$\sigma$ noise of  
{\bf F93a} (dotted line).}
\label{f:residual_CH}
\end{figure}

\begin{table}[h]
\begin{tabular}{|c|ccc|}
\hline
              &  \multicolumn{3}{|c|}{\CHl}                          \\                  
              &   $v_{LSR}$ (\kms)    &  $b$ (\kms)          &  $\log_{10}N$         \\
\hline                                                                               
              &                       &                      &                       \\
              & 5.0$^{+0.5}_{-0.5}$   &2.0$_{-0.3}^{+0.4}$   &13.58$_{-0.06}^{+0.06}$ \\
{\bf 1993}    & 6.2$^{+0.8}_{-0.8}$   &2.6$_{-0.6}^{+1.5}$   &13.20$_{-0.17}^{+0.12}$ \\
              & 5.5$^{+1.3}_{-2.0}$   &5.9$_{-2.2}^{+2.5}$   &13.19$_{-0.17}^{+0.12}$ \\
              &                       &                      &                       \\
              &    5.4                & 3.0                  &  13.85                \\
              &                       &                      &                       \\
\hline                                                                               
              &                       &                      &                       \\
              & 5.6$_{-0.15}^{+0.25}$ &1.90$_{-0.09}^{+0.08}$&13.68$_{-0.01}^{+0.03}$\\
{\bf 2002}    & 7.6$_{-0.15}^{+0.25}$ &2.40$_{-0.35}^{+0.10}$&13.55$_{-0.02}^{+0.04}$\\
              & 7.0$_{-0.5}^{+0.5}$   &7.5$_{-0.7}^{+0.8} $  &13.16$_{-0.05}^{+0.04}$\\
              &                       &                      &                       \\
              &    6.4                & 3.0                  &  13.98                \\
              &                       &                      &                       \\
\hline
\hline
                         & \multicolumn{3}{|c|}{\CHPlb}          \\
\hline
           &   $v_{LSR}$ (\kms)    &  $b$ (\kms)          &  $\log_{10}N$         \\
\hline                                                                      
           &                      &                     &                       \\
           &  7.5$_{-0.2}^{+0.2}$ & 2.5$_{-0.2}^{+0.2}$ &13.58$_{-0.02}^{+0.04}$ \\
{\bf 1993} &  5.5$_{-0.2}^{+0.2}$ & 2.7$_{-0.4}^{+0.2}$ &13.41$_{-0.03}^{+0.04}$ \\
           &  --                  & --                  & --                    \\
           &                      &                     &                       \\
           &    6.7               & 3.0                 &  13.80                \\
           &                      &                     &                       \\
\hline                                                                      
           &                      &                     &                       \\
           &  7.0$_{-0.3}^{+0.3}$ & 2.2$_{-0.3}^{+0.2}$ &13.55$_{-0.03}^{+0.04}$\\
{\bf 2002} &  4.6$_{-0.3}^{+0.3}$ & 2.3$_{-0.4}^{+0.4}$ &13.30$_{-0.05}^{+0.05}$ \\
           &  7.8$_{-2.5}^{+2.5}$ & 10$_{-3}^{+4}$      &13.00$_{-0.12}^{+0.10}$  \\
           &                      &                     &                       \\
           &    6.2               & 3.0                 &  13.82                \\
           &                      &                     &                       \\
\hline
\end{tabular}
\caption{Relative velocity, Doppler parameter and column densities
derived by fitting the high resolution CH and \CHP\ 
profiles with VPFIT. Three components are assumed; results obtained with one 
single component are also given. Solutions with the lowest $\chi^2$ 
are listed but relatively different solutions cannot be rejected.
The doublet structure of the CH
ground level has been taken into account: 
the column density given in this table is the sum for the two levels.  
The reduced $\chi^2$ is always $\sim
1$.} 
\label{t:vpfit}
\end{table}

If we now fit independently the {\bf F93a} profile, a slightly
different solution is obtained.  The parameters are less accurate due
to the lower signal-to-noise ratio $-$ see \Tab{vpfit}. The reality of
profile variations is difficult to assess from comparison of the {\bf F93a}
and {\bf F02} parameters because the decomposition is not unique; we
give in \Tab{vpfit} solutions with the lowest $\chi^2$ but relatively 
different solutions cannot be rejected. We then perform a more direct
comparison of profiles and find that multiplication of the {\bf F93a} 
optical depth by 1.34 provides an acceptable fit to the {\bf F02} profile
(\Fig{residual_CH}; the slightly different spectral resolutions of the
two spectra have been taken into account). Therefore, we conclude that
the velocity distribution has remained roughly constant between 1993
and 2002 and that the observed line variation is essentially due to 
an increase by the same factor of the column density for each of the 
three components. 

For the velocity profile obtained above, the ratio ($\delta N/N$)/($\delta W_{\lambda}/W_{\lambda}$) needed to
translate \ew\ variations into $N$ variations 
increases from 
1.37 to 1.50 
between $W_{\lambda}=48$\ and $W_{\lambda}=60$\ m\AA.
For a single component profile, these ratios equal 1.36 and 1.49 
respectively, showing that in our optical depth regime, ($\delta N/N$)/($\delta W_{\lambda}/W_{\lambda}$)
depends very weakly on the exact model chosen to fit the profiles.  
This implies an increase of $\Delta N/N=22\%$ in either the step 
function fit or in the linear fit (between 1992 and 2001). 
In this latter case, bounds of $12\%<\Delta N/N<38\%$ are obtained 
at the 3$\sigma$\ level. 
The higher (34\%) increase in $N$(CH) between {\bf F93a} and {\bf F02} 
corresponds to a 12 m\AA\ increase in $W_{\lambda}$.

\subsection{Structure in \CHP}

\CHP\ transitions are observed at 3957 \AA\ and 4232 \AA\ in the OHP data;
some additional lines are detected at shorter wavelengths in G. Herbig's
spectrum. 
The absorption line at 3957 \AA\ is free of blending and the continuum can
be accurately determined.
Although the 4232 \AA\ line is more difficult to measure, the S/N ratio
in this region is higher and thus comparable accuracy is obtained for
the two lines. 

\subsubsection{Small scale variations}

Both lines appear to be roughly constant in \ew\ 
(\Fig{result_ohp}b, c). 
Erratic variations still seem to be present, although
the larger errors yield an acceptable value of $\chi^2\sim 1$
for the best fit.
In particular, the pattern characterising the time
variations is similar for both \CHP\ lines 
and for the \CHl\ transition.
This is especially clear for the last four OHP measurements, marked
with letters A to D in \Fig{result_ohp}. We also note that the 
{\bf F02} measurement, which has been made in completely independent 
conditions, apparently displays this coherent CH/\CHP\ behavior. 
Finally, it appears in \Fig{result_ohp}a and \Fig{result_ohp}b 
that closely-spaced measurements (made during winter 1999-2000) 
display relatively smaller scatter, indicating a possible cut-off 
in the structure at scales below 1 - 2 AU.

Such a coherent variation could be an artefact due to 
instrumental effects (e.g., if changes in the background 
were not taken into account properly). We searched for any 
dependence on the observing conditions (e.g., high or low sky 
background) and found none. We also selected a few narrow stellar 
lines near 4300 \AA\ for which the continuum level is well 
defined and measured their \ew; no significant variations of the kind
observed for CH or \CHP\ are observed. Finally, we checked that 
the depth of the stellar H$\gamma$ absorption is stable as it should
be in the absence of uncontroled variations of the zero level.

We therefore strongly suspect that stochastic correlated
fluctuations of $N$(CH) and $N$(\CHP) are present, although more 
accurate measurements (especially for \CHP) would be useful to 
establish more firmly their reality and to characterise their properties. 
Among measurements performed after November 1999, the
r.m.s. fluctuation of $W_{\lambda}$(4300) is  1.5 m\AA\ while that of $W_{\lambda}$(4232)
is 1.8 m\AA\ (the corresponding fluctuations in $N$ are discussed below).

\subsubsection{Large scale variations}

\Fig{chi}b shows the result of a linear fit to the whole set of data
for \CHPlb. The minimum $\chi^2$ is again close to 1. 
Although the best fit is a slow decrease, $dW_{\lambda}/dt = -0.4$  m\AA\ yr$^{-1}$, 
a constant value, $W_{\lambda}\sim 40-43$ m\AA, cannot be rejected at the
1.5$\sigma$ level. We obtain the following 3$\sigma$ bounds: 
$-1.2<dW_{\lambda}/dt < 0.45$ m\AA\ yr$^{-1}$. Clearly, the variation observed for
\CHP\ is quite different from that seen for CH.

Consistent results are obtained from \CHPla. The ratio of the two total \CHP\ 
equivalent widths, $W_{\lambda}$(3957)/$W_{\lambda}$(4232), is found to be constant
at a value of 0.65 $\pm 0.05$, indicating that these 
lines are nearly optically thin (ratio of 
0.60). The best linear fit is also decreasing, although a constant
value for both $W_{\lambda}$(3957) or $W_{\lambda}$(4232) cannot be ruled out.

\subsubsection{High resolution observations}

\begin{figure}
 \centerline{\includegraphics[width=6cm,angle=-90]{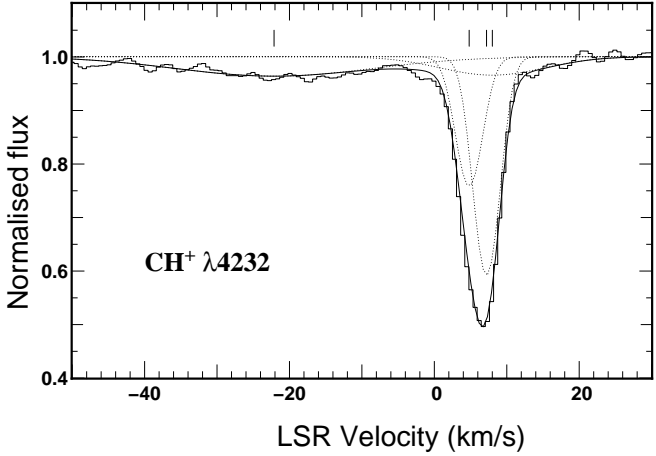}}
 \caption{High resolution observation ({\bf F02}) of the \CHP\ $\lambda$4232
 line ({\em dotted lines}: the three
   components of the fit and the stellar line; {\em solid line}:
resulting fit).}
\label{f:profCHP}
\end{figure}

\CHPlb\ has been observed twice at high resolution ({\bf F93a} 
and {\bf F02}). To fit these profiles, we first remove the blended 
stellar line using a specific Gaussian profile. 
The \CHP\ line in {\bf F02} is asymmetric, as was the \CHl\ one, 
but now it is the blue wing which is more extended. A good fit is
obtained with 3 components; parameters are summarised in \Tab{vpfit} 
and shown in \Fig{profCHP}. 

\begin{figure}
 \centerline{\includegraphics[width=8cm]{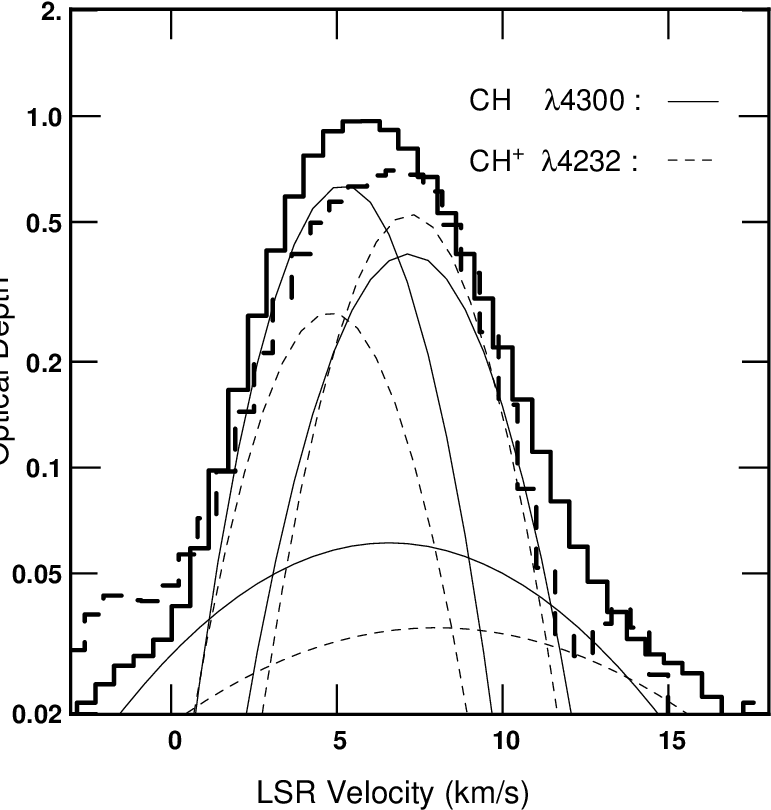}}
 \caption{Comparison of optical depth profiles for 
high resolution observations ({\bf F02}) of \CHl\ ({\em solid
line}) and \CHPlb\ ({\em dashed line}). Thick lines correspond to
observed profiles while thin ones represent individual components.}
\label{f:CHvsCHP}
\end{figure}

The {\bf F93a} and {\bf F02} profiles are almost identical. 
The same components fit both observations 
with a residual less than 1.5 and 1.0 $\sigma$, respectively. 
If the {\bf F93a} spectrum is analysed separately, a good fit 
is obtained with two narrow components only, but this is just 
due to the lower signal-to-noise ratio.
The third broad component is also apparent in the {\bf H98} 
spectrum for both the \CHP\ lines at 3957 and 4232 \AA.\\

In summary, the \CHP\ line profiles have been stable 
between 1990 and 2002. For the model derived from the {\bf F02} 
profile and the \CHPlb\ data, one gets $\delta N/N=1.3*\delta W_{\lambda}/W_{\lambda}$ 
in the range  $\delta W_{\lambda}=\ -10 {\ \rm to\ } 10$ m\AA. 
We then derive an upper limit for 
the annual variation of the column density from 1990 to 2002 of 
$-3.9<\delta N/N < 1.45\%$ per year at the 3$\sigma$ level. Tighter
constraints can be obtained by combining \CHPla\ and \CHPlb\
measurements. For all epochs, we derive $N$(\CHP) and $\sigma(N)$ 
from each transition; estimates drawn from \CHPlb\ appear to be, in
average, slightly lower than those inferred from \CHPla, presumably
because the blend with the stellar feature induces an underestimate 
of $W_{\lambda}$(4232). We thus apply a small positive and constant offset to 
$W_{\lambda}$(4232) measurements in order to get the same $<N>$ for both
transitions and finally compute the ($\sigma(N)^{-2}$ weighted)
average for each epoch. The best fit corresponds to $\delta N/N =
-0.65\%$ per year and the 3$\sigma$ bounds are $-1.5$ and $0.27\%$ per
year (a non varying $N$(\CHP) is rejected at the 2$\sigma$ level);
these values clearly exclude an increase as large as that seen for CH.

Let us now compare the velocity profiles of CH and \CHP\ absorptions.
\Fig{CHvsCHP} shows the apparent optical depth versus velocity for both
species. It can be seen that no systematic shift is present; observed 
profiles are quite similar although the asymmetry of CH and \CHP\
lines is reversed. This is clearly seen when comparing the two narrow 
components needed to fit the profiles.  (A comparison of the broad
components would be meaningless because the \CHP\ one is ill-defined
due to blending with the adjacent stellar absorption.) The CH and
\CHP\ components have the same velocities but different \CHP/CH ratios,
as often observed (see e.g., \cite{crawford89} and the $\xi$ Per
profiles given by Crane et al. 1995).

\subsection{Structure in CN}

Because CN is expected to respond strongly to density changes, the behavior
of this species can give useful clues concerning the nature of the CH
variations (Pan et al., 2001). 
The available observations were taken in 1993 ({\bf F93b}), 1998 ({\bf
H98}) and 2002 ({\bf F02}). Unfortunately, when seen towards hot stars, 
the R(0) line is
blended with stellar features from O~{\sc II} and 
C~{\sc II} (\Fig{profCN}; see Meyer \& Jura, 1985). 
As a result of the possibility of blending, we
remeasured the line in a consistent manner. In
practice, this is not easy because the resolution differs from one spectrum to
another and further, the broad stellar feature appears to vary in time.
Results are given in Table~\ref{t:result}; the larger value quoted by 
Herbig (1999) for the R(0) line is likely due to inclusion of the
stellar feature.
   
\subsubsection{Analysis of G. Herbig's spectrum}
\begin{figure}
\centerline{\includegraphics[width=6cm,angle=-90]{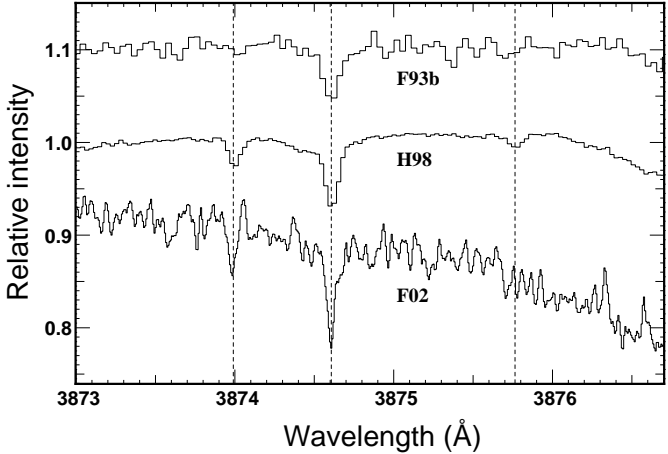}}
\caption{The CN (0,0) lines in the {\bf F93b}, {\bf H98} and {\bf F02}
 spectra. In the
{\bf H98} spectrum, the S/N ratio is 700 per 0.07 \AA\ pixel. R(0) lines
have been aligned at the rest wavelength. The three spectra have been 
normalised to 1 at $\lambda \sim 3875$ and shifted in y for
clarity. Note the broad stellar feature on the blue side of the R(0) lines.}
\label{f:profCN}
\end{figure}

Since the Keck data have the highest S/N ratio, we first discuss this
spectrum and its implication on the CN column density, velocity distribution
and excitation. The R(0), R(1) and P(1) lines are
clearly detected, as shown on \Fig{profCN}.

To our knowledge, there exist no high resolution observation with
adequate S/N of the CN lines towards HD~34078 that would allow us 
to measure their optical depth and investigate saturation effects in a
direct way. Classically, the latter are estimated by comparing the
relative strengths of the R(1) and P(1) lines (Meyer \& Jura 1985). 
However, these CN absorptions are quite weak in the HD~34078 
spectrum ($W_{\lambda}$ = 2.6 $\pm$ 0.1 m\AA\ and $W_{\lambda}$~=~1.5~$\pm$~0.1 m\AA\ 
respectively, while for the R(0) line, $W_{\lambda}$ = 5.7 $\pm$ 0.2 m\AA), and 
the \ew\ ratio poorly constrains the $b$
value. Indeed, only values smaller than 0.1 \kms\ can be rejected 
(at the 2 $\sigma$ level); thus, within errors, the \ew\ ratio is 
consistent with CN lines being optically thin. 

Another way to estimate the optical depth of the 3874.6~\AA\ R(0) 
line is to compare its \ew\ to that of the R(0) line from the 
(1,0) vibrational band at 3579.96 \AA\ (Meyer et al., 1989). 
In Herbig's spectrum, a weak feature is present at the 
expected position with $W_{\lambda}$ = 0.55 $\pm$  0.15 m\AA; in the 
optically thin limit, a value $W_{\lambda}$ = 0.50 $\pm$ 0.02 m\AA\ would 
be expected. Again, within
errors, the relative strengths of the two R(0) features are consistent
with the assumption that these two lines are thin and only a lower
limit on $b$ can be obtained: $b~>$~0.2~\kms\ (2 $\sigma$ limit).

In the optically thin limit, the observed $W_{\lambda}$ ratio for R(0) and R(1) implies 
an excitation temperature $T_{exc}$ = 3.6 $\pm$ 0.2~K (following the same
method as Meyer \& Jura, 1985) , significantly above 
the value expected $T_{exc}$ = $T_{CMB}$ = 2.73 K if the excitation 
were due only to interaction with CMB radiation. As discussed above,
the available data yield only loose constraints on $b$(CN) and if the 
latter were small enough, saturation effects might affect the
determination of $T_{exc}$. To assess the importance of these effects, 
we compute $T_{exc}$ assuming $b$ = 1, 0.5, 0.25, 0.20 \kms\ and get 
$T_{exc}$ = 3.5, 3.4, 3.0 and 2.8 K respectively. For lines of sight 
with similar 
extinction, $b$(CN) is observed to be about 1 \kms\ and values as 
low as 0.2 \kms\ appear unlikely in comparison with available 
determinations (Crawford, 1995).  
Such low values for $b$ would also be at odds with the results for 
\CHl\ ; no component in CH, that is associated with CN, has such a low 
$b$-value.  
We conclude that $T_{exc}$
very likely exceeds $T_{CMB}$, implying that 
collisions with electrons contribute significantly  
in populating the $N$ = 1 level (\cite{blackal91}). If $b$(CN)~=~1~\kms,
$T_{exc}$ would be among the highest values quoted by 
Black \& van Dishoeck (1991).

\subsubsection{McDonald 2002 and 1993 spectra}

Very recently, the CN (0,0) lines have been reobserved at McDonald (R
= 170 000) (\Fig{profCN}). The stellar feature around the R(0) line is
not well defined due to limited signal to noise; apparently, it is 
weaker than in Herbig's spectrum. The R(0) and R(1) lines are clearly 
detected with $W_{\lambda}$ = 6.3 $\pm$ 0.5 m\AA\ and $W_{\lambda}$ = 2.6 $\pm$ 0.4 m\AA\, 
respectively. These values are compatible with the 1998 ones.  On the 
other hand, the 1993 ({\bf F93b}) spectrum yields 
$W_{\lambda}$ = 5.1 $\pm$ 0.2 m\AA\ and $W_{\lambda}$ $<$ 2.0 m\AA\ (2 $\sigma$ limit) 
for the R(0) and R(1) lines, implying $T_{exc} < 3.1$ K.  
Altogether, the data suggest that both the $N$ = 0 and $N$ = 1 CN column
densities have increased, even if the variation is only marginally 
significant ($\Delta N/N = 12 \pm 6$\%). Therefore, the CN variation
could be as high as that for CH, but not much larger.

\subsection{Structure in DIBs Carriers}

\begin{figure}
	\unitlength=1cm
	\begin{picture}(10,13.5)
        \put(0,0){\psfig{width=8.5cm,angle=-90,figure=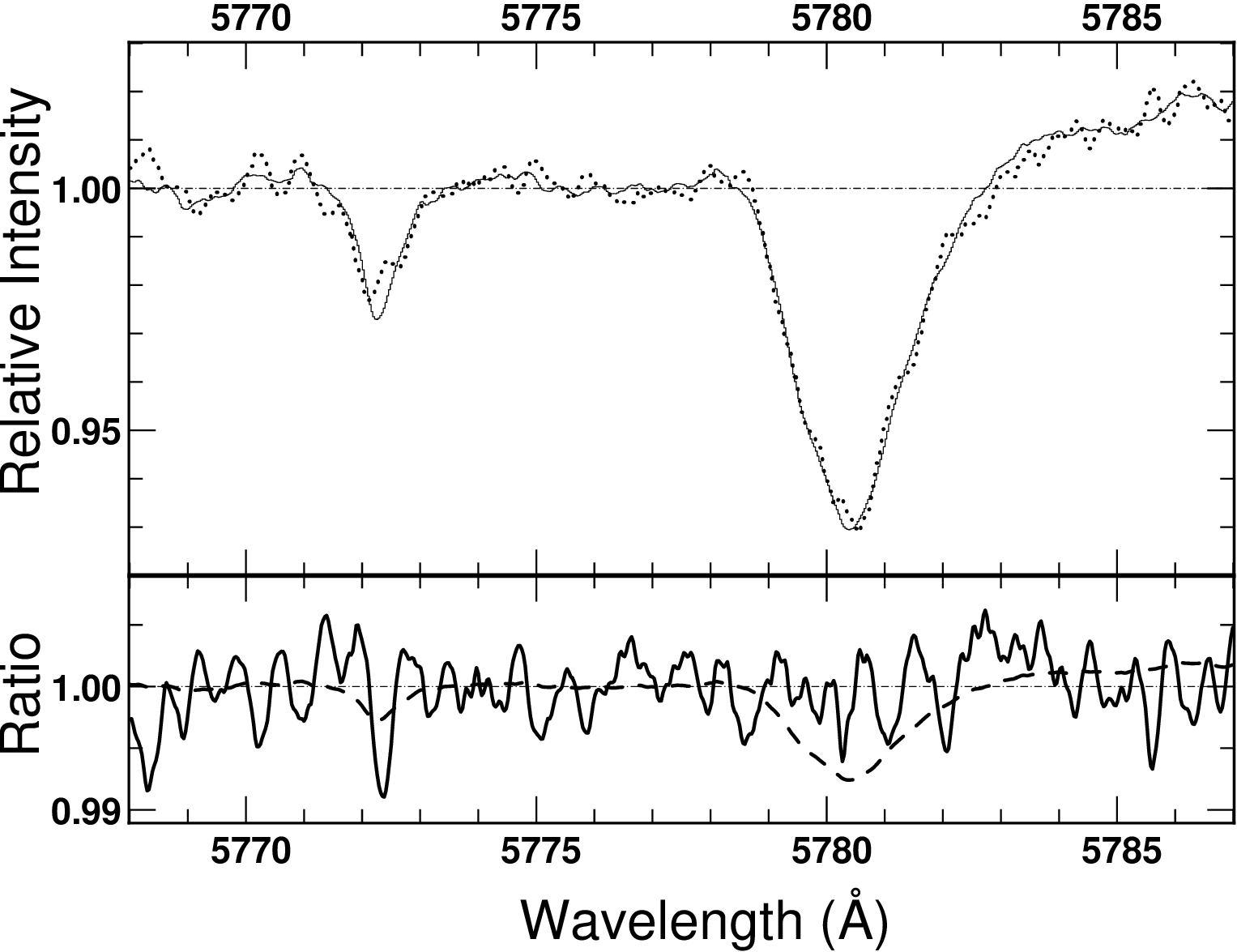}}
        \put(0,7){\psfig{width=8.5cm,angle=-90,figure=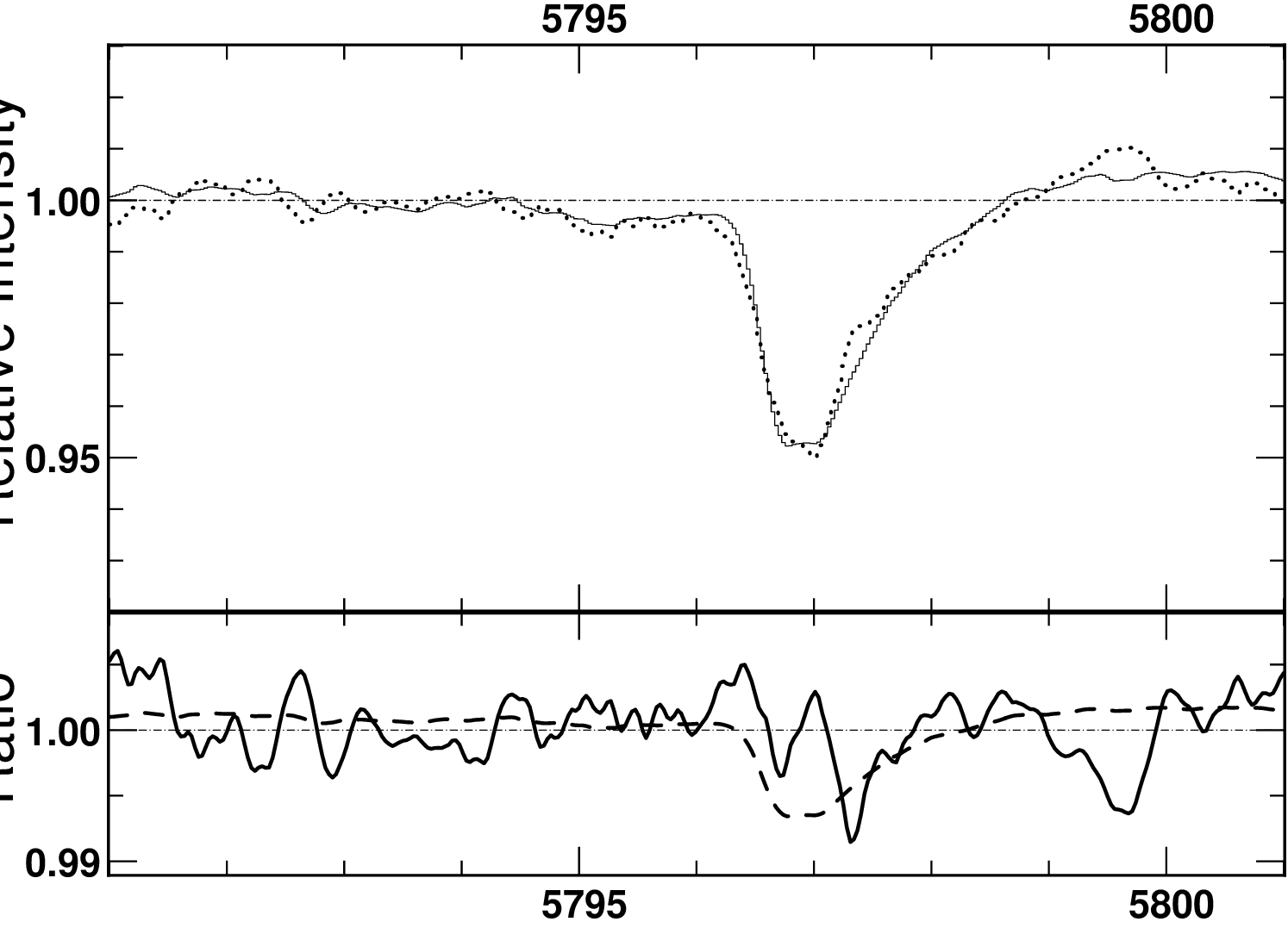}}
	\end{picture}
\caption{Time variation of the 5780 \AA\ and 5797 \AA\ DIBs. 
{\em Upper panels}: the average of all OHP spectra (solid line)
is compared to the spectrum obtained by Jenniskens et al. (1992) (dotted line).
% Note the anomalously strong DIB near 5772 \AA.
{\em Lower panels}: the ratio of the OHP and {\bf J92} spectra (solid
line) is compared to the ratio obtained after artificially increasing
by 10\% and 15\% the respective strengths of the 5780 and 5797 \AA\ DIBs 
in the OHP spectrum (dashed line).}
\label{f:DIBs}
\end{figure}

The OHP spectra acquired for our study of CH and \CHP\ contain several strong
DIBs. We now use these data together with older spectra to 
investigate the spatial distribution of DIBs carriers. Some of them
are known to behave differently with respect to the abundance of CH 
(\cite{krelowskial99}). Then, an immediate goal is to determine whether the
kind of small and large scale variations seen for CH are also present
for DIBs.

Among all detected DIBs, we initially selected those at 5780, 5797,
6196, 6284 and 6614 \AA\ which are relatively strong (opacity reaching
6\%) and narrow. For broader ones, the continuum level and shape is 
difficult to determine and only large variations could be
seen. Among the five features quoted above, we find that the sensitivity
needed to detect variations with an amplitude comparable to that of
the observed CH changes can be obtained only for the 5780 and 5797
\AA\ features. We then focus on these two DIBs. Since they are
fully resolved (except possibly for some substructure), we can perform
a direct comparison of the absorption profiles, once the spectra have
been brought to a common wavelength scale.  Equivalent width
measurements would not be appropriate due to large continuum placement 
errors. 
%It is noteworthy that the 5772 \AA\ DIB is
%stronger by a factor of about four in the HD~34078 spectrum, as 
%compared to other lines of sight (cf. Jenniskens \& D\'esert, 1994).

We first discuss variations at the scale of 0.5 - 1 year 
($\simeq$ 10 AU) among the spectra taken at OHP. Since the latter have
been taken with the same instrument, the shape of the continuum
adjacent to the DIBs considered is very stable, rendering the
intercomparison much easier. We selected pairs of spectra 
(October 2000/March 2002 and February 2001/October 2001) in 
which $W_{\lambda}$(4300)
measurements show the largest difference (about 5.3\% which
corresponds to relative variations in $N$(CH) of 7.2\%). No significant
variation is detected at the level of 10 and 15\% for the 5780 and
5797 \AA\ DIBs, respectively. To derive these upper limits, we
artificially increase/decrease the strength of the DIB considered in
the epoch 1 spectrum and then compare the latter to epoch 2 data. 
Then, the sensitivity reached is not large enough to detect variations
as faint as those displayed by $N$(CH) over a period of about one year.

To study variations at larger scales, we use the spectrum obtained by
Jenniskens et al. (1992) and compare it to the average of all our OHP
spectra. The ratio of the two smoothed spectra is computed, and to ensure a
good ``relative normalisation'', we fit values taken by this ratio in a
few windows adjacent to each DIB with a polynomial of order 1 or 2. 
We then divide the first spectrum by this relative normalisation. The
spectra thus obtained are shown in \Fig{DIBs} and show no variation of
the 5780 and 5797 \AA\ DIBs at respective levels of 10 and 20\%.  
Since both DIBs are optically thin, these upper limits 
are to be compared to the relative variation in $N$(CH) which is
estimated to be 20\% over the same period. We then conclude that the 5780 
\AA\ DIB at least has varied less than CH.

\subsection{Summary of observational results}

The main results that emerge from observations are the following:

- over the past 12 years (or scales of 100-200 AU in the foreground
cloud), the CH column density has increased by 12-38\%. A
linear form (with $dN/(Ndt)$ = 1.4\% yr$^{-1}$) or a step function 
($\Delta N/N$ = 22\% in 1998) can both fit this overall variation. Over the
same period, the \CHP\ column density displays a markedly different
behavior (a slow decrease with $dN/(Ndt)$ = -0.65\% yr$^{-1}$).

- on the scale of a few months ($\sim$ 5 AU), correlated fluctuations in 
$N$(CH) and $N$(\CHP) are strongly suggested by the data. A rough estimate of 
their amplitude over the period November 99 - March 02 can be inferred
from the r.m.s. scatter of $W_{\lambda}$ measurements:  
$\sigma$($N$(CH))/$N$(CH) $\approx$ 3.6\% and 
$\sigma$($N$(\CHP))/$N$(\CHP) $\approx$ 5.8\%. 

- CH and \CHP\ velocity profiles show no significant time variation. 

- CN observations indicate a moderate increase of $N$(CN) but a higher
excitation temperature in 1998 and 2002, as compared to 1993. The
recent \Texc\ value is significantly above that expected from
excitation by CMB radiation alone. 

- The 5780 and 5797 \AA\ DIBs have been stable and the former has
varied by no more than half that seen in CH.
%%%%%%%%%%%%%
\subsection{Discussion}

\subsubsection{AU scale structure and \CHP\ production}

We now discuss structure in the gas containing CH and \CHP\ at the 1 - 10
AU scale as probed by the homogeneous observations performed since
1999. We found indications for fluctuations of CH and \CHP\ 
absorptions: first, the scatter is relatively large and second, 
CH and \CHP\ fluctuations are apparently correlated. Nevertheless, 
it should be stressed that {\it these variations are quite small}. 
Indeed, part of the observed scatter can be attributed to measurement 
errors and the real scatter in $N$(CH) or $N$(\CHP) cannot exceed a few 
\%. This is much 
smaller than the H$_2$CO variations described by Moore \& Marscher (1995),
which attain 17\% and smaller than $N$(\HI) fluctuations seen in the 
atomic phase over similar scales.
In a simple model where most of the CH or \CHP\ gas is comprised of
identical discrete entities, the number of such ``clumps'' must exceed  
10$^3$ to get fluctuations that weak.

If the erratic variations seen since 1999 reflect the geometry of
localised regions where \CHP\ molecules are produced, then it is natural to
detect associated variations of CH since 
about 30\% of the CH molecules in a given sight line form from \CHP\, 
but the fraction approaches 100\% for gas with densities less than 
about 100 cm$^{-3}$ (\cite{gredelal02}). Presently, it is
difficult to check the consistency of this picture because
the relative amplitudes of the CH and \CHP\ fluctuations are poorly
constrained by the available data. Alternatively, in a picture where
\CHP\ is produced throughout the low density medium 
(e.g., Draine \& Katz, 1986, Pineau des For\^{e}ts et al., 1986, 
Spaans, 1995, Federman et al., 1996), the correlation
between CH and \CHP\ could be due to a physical mechanism which
induces local density enhancements for all species. For instance, 
compression in regions where chaotic flows happen to converge has 
been invoked to account for the AU-scale structure in atomic
gas (\cite{jenkinsal01}, \cite{hennebelleal99}) and could also be effective
in molecular gas. If the lifetime of these transient fragments is
small enough, such a process would lead to comparable values for 
$\delta N/N$(CH) and $\delta N/N$(\CHP); otherwise, the higher
density might result in efficient \CHP\ destruction through reactions
with \H2\ and rather produce an anticorrelation. 
One way to disentangle the above two scenarios in which small scale 
structure is either related to chemically active regions or just to pressure
fluctuations would be to study the behavior of other species whose 
chemistry is not directly linked to that of CH and \CHP.

\subsubsection{The long-term CH variation: evidence for a dense fragment~?}
  
Over larger scales (100 - 200 AU), CH and \CHP\ clearly display 
distinct behaviors. Given the number of independent measurements and
consistency among results, we consider that this difference is
well established, even if the form of the CH increase is
not well understood. In particular, we cannot exclude an abrupt
variation by 20\% occurring in 1997 - 1998.
The corresponding CN absorption has increased only modestly 
($\delta N/N$ =
12 $\pm$ 6\%) between 1993 and 1998, but the data strongly suggest a
change in excitation 
from \Texc\ $<$ 3.1 K  in 1993 to \Texc\ = 3.6 $\pm$ 0.17 K in 1998. 
The 1998 value is quite high in comparison with results quoted
by Black \& van Dishoeck (1991) for other lines of sight, implying a high 
electron density.   The fact that the ratio 
$N$(CN)/$N$(CH) has remained roughly constant
while $N$(CH) and $N$(CN) have increased, the ratio indicates that 
little change in gas density has accompanied the fluctuations 
in columns (see Pan et al., 2001).

Consider a model in which the increase in $N$(CH) by about 20\% 
over a time interval of 1 to 10 years is due to a newly intervening 
clump. $N$(CH) is known to vary linearly with $N$(\H2) with a ratio 
$N$(CH)/$N$(\H2) $\approx$ 4.0 10$^{-8}$ (Federman, 1982, Roueff, 2001),
Towards HD~34078, a value larger by a factor of four is inferred (see
Fig. 2 from Roueff, 2001). Assuming that such a ratio is still valid at the
scales considered here, one can easily compute the associated increase 
in $N$(\H2): 1.1 10$^{20}$ \units{cm}{-2}. 
Assuming an intermediate value of 50 AU for the transverse scale 
[$N$(CH) has remained at its ``high'' value at least for 4 years] and
a comparable size along the 
line of sight, the implied density for that hypothetical fragment 
is around 1.5 $\times$ 10$^5$ \units{cm}{-3}. In such a dense
fragment, very little \CHP\ is expected
to be present due to reactions with \H2, accounting for the absence
of an associated increase in $N$(\CHP). Similarly, the stability of 
the 5780 and 5797 \AA\ DIBs can be understood in this scenario since
DIBs carriers are known to lie preferentially in diffuse media. 
However, the CN/CH ratio is expected to be large at these 
densities and the relative
increase of $N$(CN) should be much larger than that of CH, 
contrary to what is observed. The density quoted above could be lowered if the 
structure were elongated along the line of sight (cf. the model proposed 
by Heiles (1997) for atomic gas), but a very large (and unrealistic!) aspect 
ratio would be needed to bring it to values typical for that kind 
of material (a few times 10$^2$ \units{cm}{-3}).  In 
particular, Federman et al. (1994) 
estimated $n_{tot}$ $\approx$ 200 \units{cm}{-3}\ from $N$(CN)/$N$(CH) and 
C$_2$ observations, at a time when $N$(CH) had its low value, but the 
column density ratio has not changed appreciably since then.  While it 
is not clear which CH component contains CN as well, the current version 
of the chemical model (Knauth et al., 2001, Pan et al., 2001) gives 
similar densities.

Another problem with this model (in addition to understanding how the
large overpressure inside the clump can be generated and maintained
long enough) is related to
the change in the excitation of CN suggested by the data. Since the 
variation in $N$(CN) has been quite limited (at most 20 to 30\%), the
higher \Texc\ value estimated for the whole gas in 1998 and 2002
implies an even larger \Texc\ in the clump if the rest of the gas has
remained at \Texc\ $\approx$ 2.7 K. Again, this points towards a large 
clump density or electron fraction. 

A last difficulty involves the observed velocity 
distribution.  The presence of an additional fragment on the 2002 line
of sight should have resulted in a new velocity component, but the
whole distribution (in particular the two narrow components) has
varied globally, without any noticeable change in shape. One could
imagine that the additional component is very narrow and does not show
up at R = 200 000, but such a feature would be atypical with respect
to profiles observed at very high resolution.

\subsubsection{Contribution from gas located near HD~34078~?}

The difficulties encountered above when attempting to account for
the CH - CN variations might be related to several
peculiarities of the HD~34078 line of sight: i) the large amount of
highly excited \H2, ii) the large CH/\H2\ and \CHP/\H2\ ratios (note
however that the CH/\CHP\ ratio is in good agreement with the
CH - \CHP\ correlation observed by Gredel (1997)),
and iii) the large \Texc\ value for CN.
%and iv) the strong DIB at 5772 \AA. 
Since point i) is attributed to the
influence of the UV bright star upon nearby material (Le Petit et
al., 2001, Boiss\'e et al., 2002), one might
wonder whether the properties observed for CH, \CHP\ and CN could also
be the result of this interaction. In this model, CH, \CHP\ and CN are
expected to be largely photodissociated in the PDR where \H2\ is 
excited to high energies. 

Then, to account for the observed 
CH and \CHP\ amounts, one may imagine that \CHP\ production occurs 
in the PDR itself, as has been suggested for the Pleiades 
(White, 1984) and IC~348 (Snow, 1993).
In this scenario, \CHP\ is the result of reactions between C$^+$ and H$_2$
in gas at elevated temperatures caused by enhanced H$_2$ photodissociation.
The large amount of CH then arises from the synthesis of \CHP and the
increase in $T_{exc}$ occurs because there is enhanced ionization 
produced by the star's UV radiation. One potential difficulty involves
the velocity distribution which appears to be quite narrow (except for
the broad component) and stable (cf. Herbig 1999). This 
suggests quiescent absorbing material rather than gas located near a
bow shock around a O9 star moving at more than 100 \kms. 
Possibly, we are seeing "composite absorption" from both pieces of gas,
one located close to HD~34078 and strongly influenced by it, and 
the other consisting of more
distant material where the radiation field is enhanced by only a
factor of a few relative to the average galactic field. 

\subsubsection{Prospects}
 
Let us now suggest a few prospects
concerning additional observations that might clarify the situation.
First, we wish to stress the importance of data other than those
considered in this paper that could have been taken
before 1999, especially spectra obtained in 1997 - 1998,
which would be very useful
in constraining the shape of the long-term variation in CH. We thus prompt
any reader who would have such data in hand to contact us.
Evidently, it is important to continue the CH and \CHP\ observations with
increased S/N ratio in order to characterise better the properties of
erratic fluctuations on timescales of about one year. At the same
time, these data would give information on the shape of the long-term
variation of CH and \CHP\ absorptions.

To date, FUSE observations have provided 4 spectra, all taken after 1999;
their analysis indicates that $N$(\H2) variations - if any - are smaller
than
about 5\% (Boiss\'e et al., 2002). If $N$(CH) keeps increasing, we shall
be able to verify whether the CH - \H2\ correlation still holds at the 10
-100 AU scales since the expected increase in $N$(\H2) will exceed our
detection limit. The behavior of other species like C~{\sc I} and CO can be
investigated also using these spectra. However, this can be done with
better sensitivity with HST thanks to higher S/N and
spectral resolution. HST/STIS spectra would provide invaluable
information  on the density via C~{\sc I} fine
structure transitions and C$_2$ lines and allow us to follow with improved
accuracy the small and large scale variations of many species
(C~{\sc I}, H~{\sc I}, excited H$_2$, CO, C$_2$...). Repeating the C$_2$
visible observations reported by Federman et al. (1994) could also give
direct evidence for the high clump density quoted above.

%%%%%%%%%%%%%%%%%%%%%%%%%%%%%%%%%%%%%%%%%%%%%%%%%%%%%%%%%%%%%%%%%%
\acknowledgements{We thank the astronomers and staff who obtained the OHP
spectra, M. Allen, F.X. D\'esert, G. Herbig, J. Krelowski, G. Galazutdinov
and F. Musaev who kindly provided their data, and
E. Roueff and M. Gerin for obtaining an IRAM CO spectrum.  We acknowledge
fruitful discussions with J. Black, P. Hennebelle and G. Pineau des 
For\^ets  and thank S.
Thoraval for his initial participation in this program.  S.R.F. was
supported in part by NASA LTSA grant NAG5-4957.}

%%%%%%%%%%%%%%%%%%%%%%

%======================================
\newpage
\onecolumn
\appendix

\section{Tables...}
\begin{table}[htb]
 \begin{tabular}{|l|c|c|c|c|cl|}
 \hline
 Date & Observatory & S/N & R$^a$ & Transition$^b$ & References& \\
 \hline
 9-89 $->$1-91&KPNO$^c$& 135 &85 000&CH, \CHP&Allen, 94 &{\bf A91}\\
\hline
06-12/1991&OHP (Aur\'elie)&$>$100&14 300&CH, \CHP(4232), DIBs&Jenniskens et al., 92 &
{\bf J91}\\
\hline
11-14-97 &  Terskol& 50-80 & 45 000&CH, \CHP& Krelowski et al., 99& {\bf K97a}\\
11-19-97  & Terskol& 50-80 & 45 000&CH, \CHP& Krelowski et al., 99& {\bf K97b}\\
 \hline 
 01-93& McDonald echelle   & 45-65 & 200 000  & CH, \CHP(4232)& & {\bf F93a}\\
 12-93& McDonald Sandiford & 100-150  & 60 000 & CH, \CHP, CN& & {\bf F93b}\\
 02-02& McDonald echelle   & 100-150 & 170 000 & CH, \CHP(4232), CN& & {\bf F02}\\
\hline
 10-98&Keck&600&45 000&CH, \CHP, CN&Herbig, 99 & {\bf H98}\\
\hline

 11-18-99& &135-140& &&& {\bf R99a}\\
 12-22-99& &310-360& &&& {\bf R99b}\\
 1-27-00 & &120-200& &&& {\bf R00a}\\
 2-27-00&OHP (Elodie) &105-115&32 000 &CH, \CHP, DIBs && {\bf R00b}\\
 3-18-00& &125-150& &&& {\bf R00c}\\
 10-01-00& &120-155& &&& {\bf R00d}\\
 2-10-01& &105-145& &&& {\bf R01a}\\
 10-2-01& &105-145& &&& {\bf R01b}\\
 3-4-02& &150-160& &&& {\bf R02}\\
\hline
\end{tabular}
\caption{List of Observations. $^a$\ Resolution near CH$\ \lambda
4300$. $^b$\ Interstellar lines that have been 
included in this study. CH stands for
CH$\ \lambda 4300$; CH$^+$ for CH$^+\ \lambda 3957$ and CH$^+\ \lambda 4232$, 
unless specified;   CN for CN
$\lambda 3875$  (Table~\ref{t:result}). DIBs stands   
for diffuse interstellar bands (see text for details).}
\label{t:observation}
\end{table}

\begin{table}[htb]
 \begin{tabular}{cllllllll}
 \hline
             &CH $^a$           & CH+          & CH+$^b$      & CN$^b$      & CN$^b$       \\
             &4300 \AA          &3957 \AA      & 4232 \AA         & R(0)        & R(1)         \\
\hline             
{\bf A91}    & 48.3$^{0.6}$     & 29.8$^{1.4}$ & 44.4$^{2.0}$     &  ...        & ...         \\
{\bf J91}    & 42.0$^{10.0}$    &  ...         & 52.0$^{10.0}$(b) & ...         & ...         \\
{\bf F93a}   & 48.0$^{2.0}$     &  ...         & 44.0$^{1.7}$     &  ...        & ...         \\
{\bf F93b}   & 52.0$^{1.2}$     & 29.3$^{0.8}$ & 43.1$^{0.9}$     & 5.1$^{0.2}$ & $\le$2. (c)   \\
{\bf K97a}   & 49.6$^{1.0}$     & 28.7$^{1.5}$ & 39.8$^{3.0}$     & ...         & ...         \\
{\bf K97b}   & 52.6$^{2.0}$     & 25.4$^{2.0}$ & 44.0$^{3.0}$     & ...         & ...         \\
{\bf H98}    & 58.0$^{0.7}$     & 27.3$^{0.2}$ & 42.0$^{2.0}$     & 5.7$^{0.2}$ & 2.6$^{0.1}$ \\
{\bf R99a}   & 55.7$^{0.7}$     & 29.0$^{2.2}$ & 38.0$^{2.0}$     &             &             \\
{\bf R99b}   & 55.4$^{0.5}$     & 27.9$^{1.4}$ & 41.5$^{2.0}$     &             &             \\
{\bf R00a}   & 55.2$^{0.7}$     & 26.3$^{1.5}$ & 40.3$^{2.0}$     &             &             \\
{\bf R00b}   & 57.8$^{1.2}$     & 22.8$^{2.8}$ & 40.5$^{3.0}$     &             &             \\
{\bf R00c}   & 55.8$^{1.0}$     & 24.7$^{2.3}$ & 40.0$^{2.0}$     &             &             \\
{\bf R00d}   & 56.6$^{1.1}$     & 23.2$^{1.9}$ & 39.0$^{2.0}$     &             &             \\
{\bf R01a}   & 58.8$^{0.8}$     & 31.9$^{2.3}$ & 44.0$^{3.0}$     &             &             \\
{\bf R01b}   & 55.4$^{0.7}$     & 24.4$^{1.3}$ & 39.0$^{3.0}$     &             &             \\
{\bf F02}    & 60.0$^{1.0}$     & ...          & 44.0$^{2.0}$     & 6.3$^{0.2}$ & 2.6$^{0.4}$ \\
{\bf R02}    & 58.1$^{1.0}$     & 25.3$^{1.6}$ & 42.0$^{2.0}$     &             &             \\
\hline             
\end{tabular}
\caption{Equivalent width of interstellar lines (all errors are 1$\sigma$ error). 
$^a$\ Includes broad absorption (cannot recover the broad component).
$^b$\ Does not include broad absorption from stellar line 
(visible in all spectra).
$^c$\ 2$\sigma$ upper limit.}
\label{t:result}
\end{table}

\end{document}